\documentclass[graybox]{svmult}

\usepackage{mathptmx}       
\usepackage{helvet}         
\usepackage{courier}        
\usepackage{type1cm}        
\usepackage{makeidx}         
\usepackage{graphicx}        
\usepackage{multicol}        
\usepackage[bottom]{footmisc}

\makeindex             

\begin{document}

\title*{Attention on Weak Ties in Social and Communication Networks}
\author{Lilian Weng, M\'arton Karsai, Nicola Perra, Filippo Menczer and Alessandro Flammini}
\institute{Lilian Weng \at Affirm Inc., \email{lilian.wengweng@gmail.com}
\and M\'arton Karsai \at Laboratoire de l'Informatique du Parall\'elisme,
	INRIA-UMR 5668, IXXI,  ENS, Lyon, France \email{marton.karsai@ens-lyon.fr}
\and Nicola Perra \at Centre for Business Networks Analysis, University of Greenwich, London, UK \email{n.perra@greenwich.ac.uk}	
\and Filippo Menczer \at Center for Complex Networks and Systems Research, School of Informatics and Computing, Indiana University, Bloomington, Indiana, USA \email{fil@indiana.edu}
\and Alessandro Flammini \at Center for Complex Networks and Systems Research, School of Informatics and Computing, Indiana University, Bloomington, Indiana, USA \email{aflammin@indiana.edu}}
%
%
\maketitle

\abstract{Granovetter's weak tie theory of social networks is built around two central hypotheses. The first states that strong social ties carry the large majority of interaction events; the second maintains that weak social ties, although less active, are often relevant for the exchange of especially important information (e.g., about potential new jobs in Granovetter's work).  While several empirical studies have provided support for the first hypothesis, the second has been the object of far less scrutiny. A possible reason is that it involves notions relative to the nature and importance of the information that are hard to quantify and measure, especially in large scale studies. Here, we search for empirical validation of both  Granovetter's hypotheses. We find clear empirical support for the first. We also provide empirical evidence and a quantitative interpretation for the second. We show that attention, measured as the fraction of interactions devoted to a particular social connection, is high on weak ties --- possibly reflecting the postulated informational purposes of such ties --- but also on very strong ties. Data from online social media and mobile communication reveal network-dependent mixtures of these two effects on the basis of a platform's typical usage. Our results establish a clear relationships between attention, importance, and strength of social links, and could lead to improved algorithms to prioritize social media content.}

\section{Introduction}
With the aid of Internet technologies we can easily communicate with essentially anybody in the world at any time. Social media platforms, for example, provide inexpensive opportunities of creating and maintaining social connections and of broadcasting and gathering information through these connections~\cite{weng2013shortcut}. In fact, the huge amount of information that we create and exchange exceeds our capacity to consume it~\cite{dunbar_social_1998,goncalves_validation_2011} and increases the competition among ideas for our collective attention~\cite{backstrom2011center,weng2012scirep,Hodas2012_attention}.  As a result, our interactions are steered more than ever before by the ``economy of attention''~\cite{simon71,davenport01}.  As Simon predicted: 
\begin{quote}
``What information consumes is rather obvious: it consumes the attention of its recipients. Hence a wealth of information creates a poverty of attention and a need to allocate that attention efficiently among the overabundance of information sources that might consume it."~\cite{simon71}
\end{quote}
Attention has thus become a valuable resource to be spent parsimoniously. Here we investigate how individuals allocate attention to different classes of social connections.

In the seminal paper ``The strength of weak ties,'' Granovetter~\cite{Granovetter} defines the \emph{strength} of social ties as proportional to the size of the shared social circles of connected individuals. The more common friends two individuals have, the stronger is the tie between them.  We adopt this same definition here. 
In the \emph{weak tie hypothesis}, he postulates that social ties of different strength play distinct roles in the dynamics of social structure and information sharing~\cite{Granovetter,granovetter1995book}. In particular, weak ties do not carry as much communication as strong ties do, but they often act as bridges between communities, and thus as important channels for novel information otherwise unavailable in close social circles.

There is a vast literature supporting the idea that weak ties play an important role in spreading novel information across communities~\cite{Brown:1987kx,levin2004strength,JP2007pnas,gilbert2009predict,bakshy2012role}. 
This body of work, however, is not concerned with the nature and importance of the information exchanged across ties, and in particular does not confirm (or disprove) the second of Granovetter's hypotheses, namely that weak ties carry ``important'' information. One major aim of this chapter is to address this second, more subtle, point by measuring the \emph{attention} that users pay to information exchanged on ties of different strength.

Specifically, here we address two questions:

\begin{enumerate} 
\item How is the \emph{intensity} of communication related to the strength of a social tie?
\item How is \emph{attention} differently allocated among strong and weak ties?
\end{enumerate}
Answering these two questions leads us to naturally discriminate between ties of different strength and the kind of interactions they represent. In particular we study how social exchange and information gathering interactions are typically related to the strength of the ties.
We investigate these questions using three large-scale networks describing different types of human interactions: information sharing in online social media, cell phone calls, and email exchanges.

The first question can be quantitatively addressed by measuring the \emph{strength} of a social tie as the size of the neighborhood shared by two connected agents. Our results, in agreement with previous studies (e.g., by Onnela et~al.~\cite{JP2007pnas}), confirm the first of the weak tie hypothesis: the largest fraction of interactions do happen on strong ties while weak ties carry much less traffic~\cite{Granovetter,JP2007pnas}.
We then focus on the second of Granovetter's hypotheses by examining the role of \emph{attention} and its relationship with tie strength. We propose to use attention as a proxy for the importance of the information exchanged across a tie. Attention is here defined as the fraction of an individual's activities that is devoted to a particular tie. We study how attention changes as a function of the strength of ties, and examine how it is distributed among the user's ties to either access information or maintain social connections. Interestingly, we find that only very weak or very strong ties attract a good amount of attention, implying two potentially competing trends. On one hand, people frequently interact with strong ties to satisfy their social needs. On the other hand, people look for information through weak ties, as suggested by both Granovetter's and Simon's work. The former activity assigns more attention to strong ties, while the latter prefers weak ones. While these observations hold across all the datasets we examine, the relative magnitude of the two tendencies depends on the specific network functionality.

\section{Related Work}

Motivated by Granovetter's work, many empirical studies explored the role of weak ties in social networks mostly by surveys or interviews, and found support for the weak tie hypothesis~\cite{friedkin1980test,lin1981social,granovetter1983,Brown:1987kx,nelson1989strength,levin2004strength}. 
Brown and Reingen~\cite{Brown:1987kx} found an important bridging function of weak ties in word-of-month referral behavior, allowing information to travel from one distinct subgroup of referral actors to another. 
Levin and Cross~\cite{levin2004strength} investigated dyadic social ties in transferring useful knowledge. They found that strong ties lead to the reception of useful knowledge more than weak ties, but weak ties benefit knowledge transmission when the trustworthiness is controlled.
Gilbert and Karahalios~\cite{gilbert2009predict} tested several dimensions of tie strength on social media and revealed that both intensity of communication and intimate language are strong indicators of relationship closeness. 
Strong ties are also believed to provide greater emotional support~\cite{haythornthwaite,wellman1990different} and to be more influential~\cite{Brown:1987kx,bakshy2012role,bond2012nature}, while weak ties provide novel information and connect us to opportunities outside our immediate circles~\cite{Granovetter,putnam2001bowling,burt2009structural}.

Advances in technology have lowered the cost of communication, information production and consumption, and social link formation, creating unprecedented opportunities to study social interactions through massive digital traces~\cite{lazer2009,Vespignani2009}. 
However, only a handful of studies have leveraged recently available large-scale data to explore the weak tie hypothesis. 
Onnela et~al.~\cite{JP2007pnas} analyzed a mobile call network and showed that individuals in clusters tend to communicate more, while weak ties, acting as bridges between clusters, have less traffic. Bakshy et~al.~\cite{bakshy2012role} found that on Facebook, strong ties are individually more influential in propagating information (external URLs) compared to weak ties. However, the greater number of weak ties collectively contribute to a larger influence in aggregate~\cite{Demeo2014_Facebook}.  
Weak ties also play a dominant role in slowing down information spreading in temporal networks, due to their special topological bottleneck position and limited communication frequencies~\cite{karsai2011small,Miritello2011,karsai13-1,ubaldi2015asymptotic}. The presence of strong and weak ties has been recently linked also to the opposite effect. In fact, the concentration of interactions between strong ties facilitates classes of contagion processes characterized by endemic states such as Susceptible-Infected-Susceptible (SIS) processes~\cite{sun2015contrasting}.

The body of empirical work referenced above includes both small experiments conducted in controlled settings and ``big data'' approaches. As an introduction to the work presented here, it is important to stress the different advantages that these two approaches  bring to the study of weak and strong ties. Big data approaches have obviously the advantage of scale, and, often, of addressing questions in the wild. Their major weakness is that they provide much less control on the nature of specific social ties and of information exchanged. Here we try to overcome this limitation by adopting attention as a proxy for the importance of the information exchanged and as a tool to infer the nature of a tie. 

\section{Datasets and Network Representation}

We consider three very different datasets. The basic statistics of each network are summarized in Table~\ref{table:dataset}. 

\begin{table}
	\caption{Statistics of three network datasets. Note that a link $(i,j)$ is deemed mutual if both $(i,j)$ and $(j,i)$ exist in the network.}
	\begin{tabular}{r|r|r|r|r|r}
		\hline
		Network name & \# Nodes & \# Links & \% Mutual links & Weight & Duration \\
		\hline
		Twitter & 628,916 & 44,611,893 & 64\% & \# reposts & Mar--Apr 2012 \\
		Cell phones & 6,101,641 & 19,013,221 & 61\% & \# calls & 120 days \\
		Email & 86,818 & 359,817 & 16\% & \# messages & Sep 1999--Feb 2002 \\
		\hline
	\end{tabular}
\label{table:dataset}
\end{table}

\begin{description}

\item [Twitter network.] Twitter is a micro-blogging platform used by many millions of people to broadcast short messages through social connections. Users can subscribe to (or ``follow'')  people they deem interesting to automatically receive the information they produce. The collection of all ``follow" connections forms the \emph{follower network.} In the follower network, each node $i \in V$ represents a user and a directed link $(i,j) \in E$ is drawn between nodes $i$ and $j$ if user $i$ follows $j$. In such a directed link, we call $i$ the \emph{source} node and $j$ the \emph{target} (but note that information travels in the opposite direction). Users post short messages (``tweets''), which may be reposted (``retweeted'') by their followers. We define the weight of a link $(i,j)$ as the number of times that $i$ retweets $j$.

Twitter allows for other forms of interaction, such as direct mentions of specific users. While these could alternatively be used to define edge weights, mentions are typically used in discussions and do not necessarily indicate replies to previous tweets. Retweets provide a more direct measure of the extent to which a user $i$ pays attention to information broadcast by $j$.

We collected about 934 millions tweets, 150 millions of which were retweets, from a 10\% sample of the public tweets provided by the Twitter streaming API.\footnote{dev.twitter.com/docs/streaming-apis} The information about following connections is gathered for a randomly sampled subset of creators of the collected tweets through the Twitter follower API.\footnote{dev.twitter.com/docs/api/1/get/followers/ids}

\item [Phone call network.] The mobile phone call dataset records about 487 millions call events during 120 days with one second resolution. The dataset was recorded by a single operator with $20\%$ market share in an undisclosed European country.\footnote{A statement about the ethical use of this dataset was issued by 
Northeastern University's 
Institutional Review Board.} This dataset naturally leads to a social network where nodes represent users, and a direct edge $(i,j) \in E$ is present if target user $j$ has received at least one call from source user $i$. The weight of each tie represents the number of calls.

\item [Enron email network.] The Enron email network records 246,391 emails exchanged inside the Enron corporation. An edge $(i,j) \in E$ is established if there is at least one email from source user $i$ to target user $j$, as $i$ directs individual attention to $j$ intentionally. The weight of an edge is the number of emails from $i$ to $j$. The Enron email corpus was made publicly available during the legal investigation concerning the Enron corporation~\cite{klimt2004enron}.
\end{description}

\section{Tie Strength, Weight, and Attention}

\subsection{Tie strength}

In line with Granovetter's hypothesis, we measure \emph{tie strength} --- the closeness between two connected users $i$ and $j$ --- as the Jaccard coefficient between their friend sets~\cite{Granovetter,JP2007pnas}:
\begin{equation}
O_{ij} = \frac{|N_i \cap N_j|}{|N_i \cup N_j  \setminus \{ i, j\} |}
\end{equation}

where $N_i$ and $N_j$ are the sets of neighbors of $i$ and $j$, respectively:
\begin{equation}
N_i = \{u \; | \; (i,u) \in E \lor (u,i) \in E \}.
\end{equation}

In measuring the strength of a tie according to this definition, we ignore the direction of links. Although considering direction might convey a more nuanced interpretation of the notion of strength itself, it would require introducing an additional hypothesis not directly testable, which we prefer to avoid in this study.
Link direction is obviously important when one is concerned with the flow of information, therefore we will consider it later when we examine the information and attention flows.

\begin{figure}
\centering
 \includegraphics[scale=0.4]{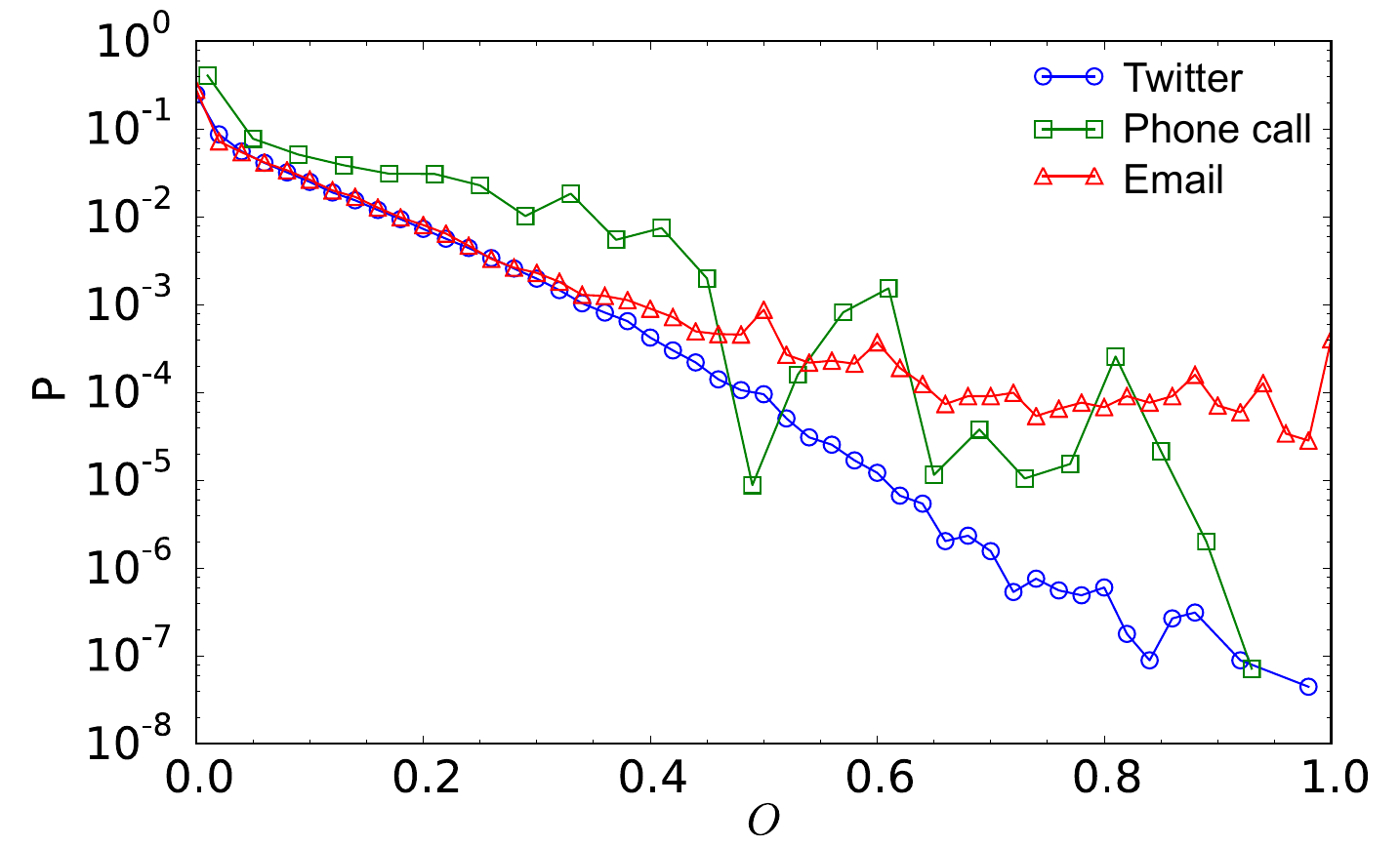}
\caption{Distribution of link overlap. We plot the probability distributions of link overlap for the three datasets. }
\label{fig:distr_overlap}
\end{figure}

In the subsequent discussion we also refer to tie strength as \emph{link overlap.} In Fig.~\ref{fig:distr_overlap} we plot the probability distribution of link overlap in the three datasets. All of them present fast (exponential) decay: most ties are weak with little overlap, while only a very small fraction of ties are strong.

\begin{figure}
 \includegraphics[width=\columnwidth]{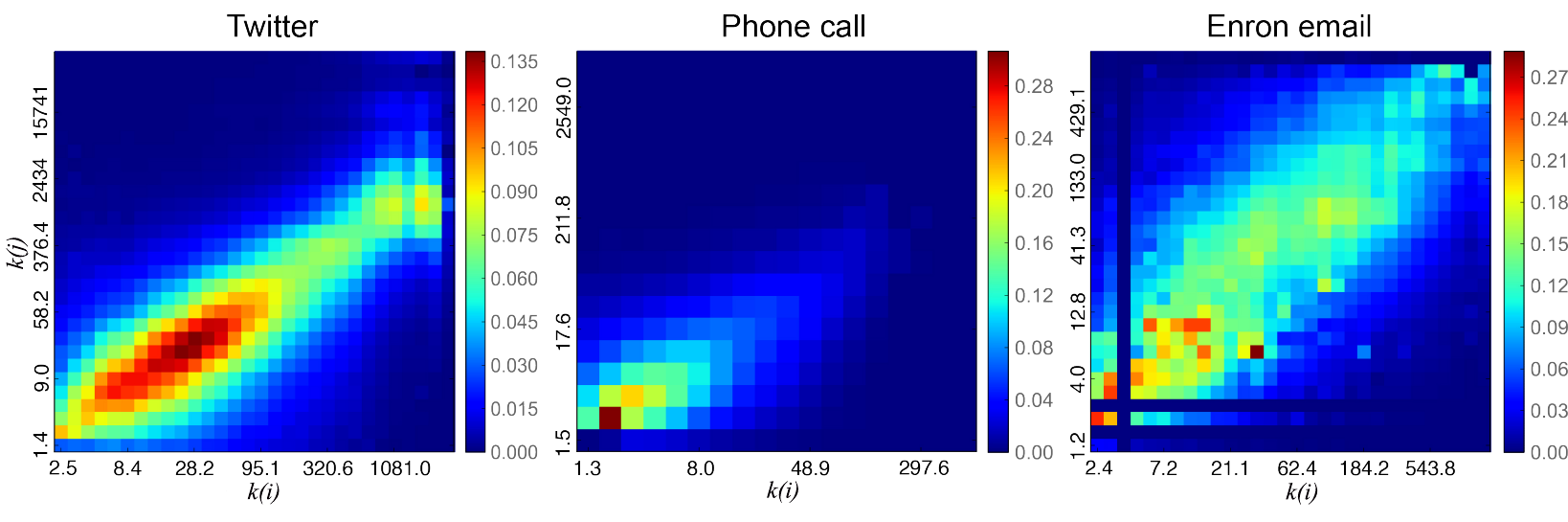}
\caption{Tie strength as a function of the degree. Heat maps of link overlap of an edge $(i,j)$ as a function of degree $k(i)$ of the source node $i$ and degree $k(j)$ of the target node $j$ in Twitter, cell phone network and Enron email network. Degrees are plotted using logarithmic bins. The color of each cell represents the average link overlap of all the edges that fall into that bin given the degrees of the target and source nodes.
Note that the degree is the sum of in-degree and out-degree, i.e. the number of neighbors of a given node irrespective of direction.}
\label{fig:heatmap_ov}
\end{figure}

The heat maps in Fig.~\ref{fig:heatmap_ov} show tie strength as a function of the degrees of the two nodes connected by the link. In Twitter, high link overlap is more likely to appear between two nodes with similar degrees; in the cell phone call network, ties between users with fewer contacts tend to have higher overlap; in the Enron network, people with similar numbers of email contacts are more likely to have overlapping contact groups.

\subsection{Weight}

\begin{figure}
\centering
 \includegraphics[scale=0.4]{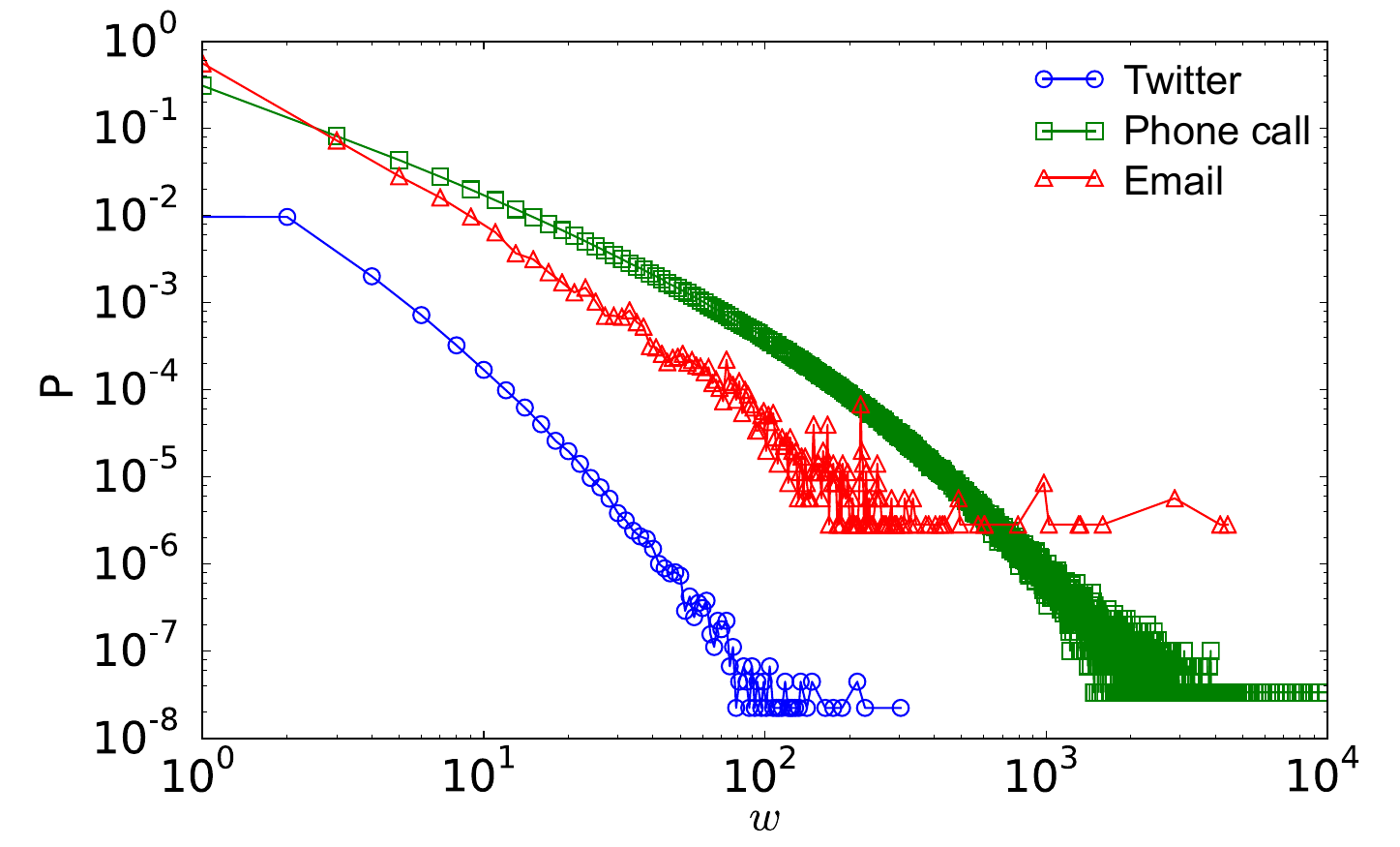}
\caption{Distribution of link weight. We plot the probability distributions of link weights for the three datasets.}
\label{fig:distr_weight}
\end{figure}

The intensity of communication on a tie $(i,j)$ is quantified by the total number of times that $i$ retweets, calls, or emails $j$, denoted as link \emph{weight} $w_{ij}$. Fig.~\ref{fig:distr_weight} shows broad distributions of link weights, suggesting that in all three networks, the majority of links carries little traffic but a significant minority supports extremely high volumes of interactions.

\subsection{Attention}

As we mentioned earlier, we propose to use \emph{attention} toward a social contact as a proxy for the importance of information provided by that contact. Attention is therefore a key notion in the present analysis. In principle we would like to have a quantity that measures the amount of cognitive resources that an individual invests in interacting with other individuals. A good proxy could be time spent on the specific ``platform" but this information is not available in our data. A second alternative would be the activity of the users (e.g., the tweets produced) but this could yield an artificially low value for users who mostly consume information. A third, computationally convenient alternative is to link attention to the number of friends a user has in the social network. It is reasonable to expect that the cognitive resources spent in maintaining social relationships is, on average, an increasing function of the degree of a node, up until a saturation limit compatible with the finite attention of individuals~\cite{dunbar_social_1998, goncalves_validation_2011,Hodas2012_attention,weng2012scirep,miritello2013time,stiller2007perspective,baronchelli2013networks,arnaboldi2013dynamics}, and after which attention should remain essentially constant. There is a considerable amount of empirical work that supports this hypothesis. Romero~et~al.~\cite{Romero2011} showed that the probability of adopting (and therefore paying attention to) a hashtag exhibits this qualitative behavior when plotted vs. the number of times the user is exposed to the hashtag --- and therefore, on average, the number of friends. 
Hodas and Lerman \cite{Hodas2012_attention} found an analogous result for the probability of retweeting a URL. Kwak et~al.~\cite{Kwak2010} observed the same qualitative behavior between user activity and both number of followers and friends on Twitter. These studies together suggest that different proxies of attention behave in a qualitatively similar fashion when considered as functions of the the degree of the user, i.e., a relatively quick growth for small values of the degree, followed by a saturation or a very slow growth regime. 

\begin{figure}
\centering
 \includegraphics[scale=0.3]{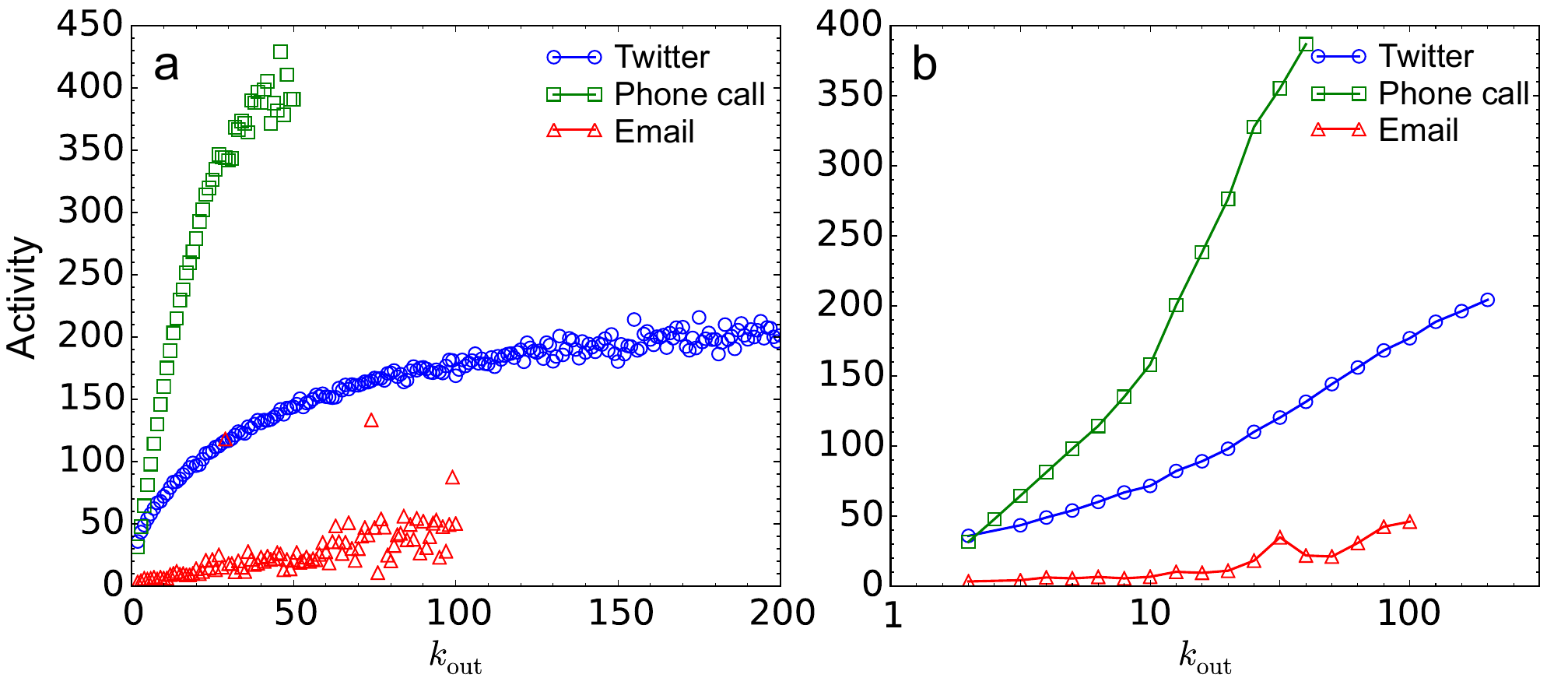}
\caption{Average activity (the number of tweets, calls, or emails) of individuals with a given out-degree on (a) linear and (b) logarithmic scales in three networks. 
We track users with up to $k_{out} = 200$ in the Twitter network, $k_{out} = 50$ in the phone call network, and $k_{out} = 100$ in the Enron email network to avoid the noise caused by scarcity of data points. More than 92\% of users in the Twitter network have $k_{out} \leq 200$, more than 99\% of users in the phone call network have $k_{out} \leq 50$, and more than 92\% of users in the email network have $k_{out} \leq 100$.}
\label{fig:kout_vs_activity}
\end{figure}

We find support for this general behavior in our datasets as well. Indeed, Fig.~\ref{fig:kout_vs_activity}(a) illustrates how activity (tweets, phone calls, emails)  grows as a function of out-degree (people one follows, calls, or emails). In general, we observe that the activity of an individual grows nonlinearly with out-degree; it can be approximated by a linear dependence in logarithmic scale (Fig.~\ref{fig:kout_vs_activity}(b)).

To capture this qualitative behavior we define the total attention of user $i$ as
\begin{equation}
a(i) = \alpha \log k_\mathrm{out}(i)
\end{equation}
and without loss of generality, we set $\alpha=1$.

\begin{figure}
\centering
 \includegraphics[scale=0.4]{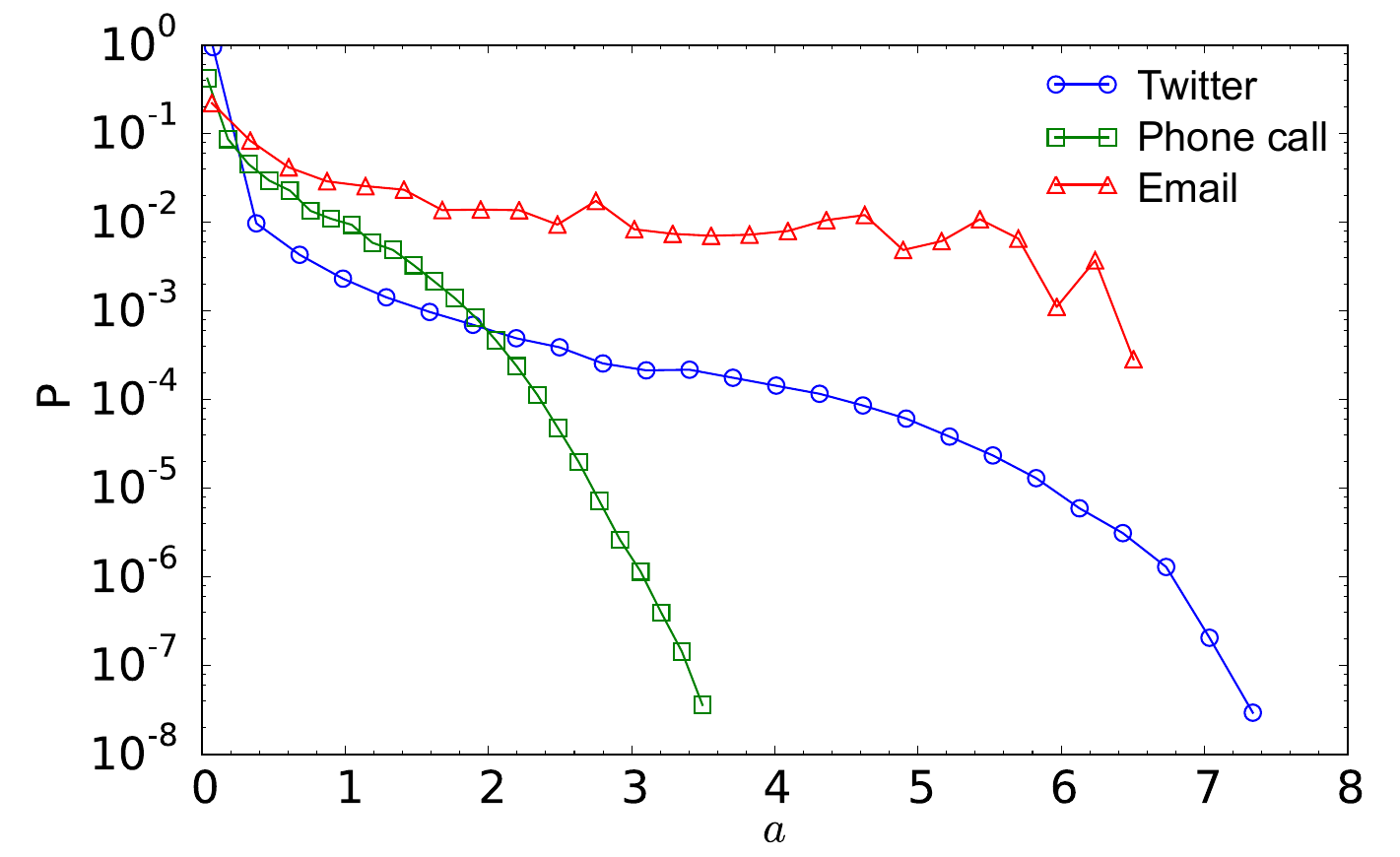}
\caption{Distribution of link attention. We plot the probability distributions of link attention for the three datasets.}
\label{fig:distr_attention}
\end{figure}

Next, we assume that the fraction of attention devoted by user $i$ to user $j$, $a_{ij}$ is proportional to the weight $w_{ij}$ of link $(i,j)$. We thus obtain:
\begin{equation}
a_{ij} = a(i) \cdot \frac{w_{ij}}{\sum_{u \in N_i^\mathrm{out}} w_{iu}} = \log k_\mathrm{out}(i) \cdot \frac{w_{ij}}{\sum_{u \in N_i^\mathrm{out}} w_{iu}}
\end{equation}
where $N_i^\mathrm{out} = \{u \; | \; (i,u) \in E \}$.
Unlike tie strength, attention considers direction, because it flows from the source to the target and only depends on the actions of the source. Attention has a narrow distribution in all three datasets, as shown in Fig.~\ref{fig:distr_attention}.

\section{Weak Ties Hypothesis and the Role of Attention}

The weak tie hypothesis maintains that strong ties carry the  majority of interactions, while weak ties act as bridges between communities and are crucial channels for transferring important or novel information. If this is true, we expect that users pay more attention to information received through a weak tie. 
In the present section we test such hypothesis by measuring how attention is allocated across strong and weak ties. 
The use of attention as a proxy for importance allows us to overcome the difficulty of defining and empirically measuring the elusive notions of importance or novelty of a piece of information.

\subsection{Traffic on Strong Ties}

\begin{figure}
\centering
 \includegraphics[scale=0.2]{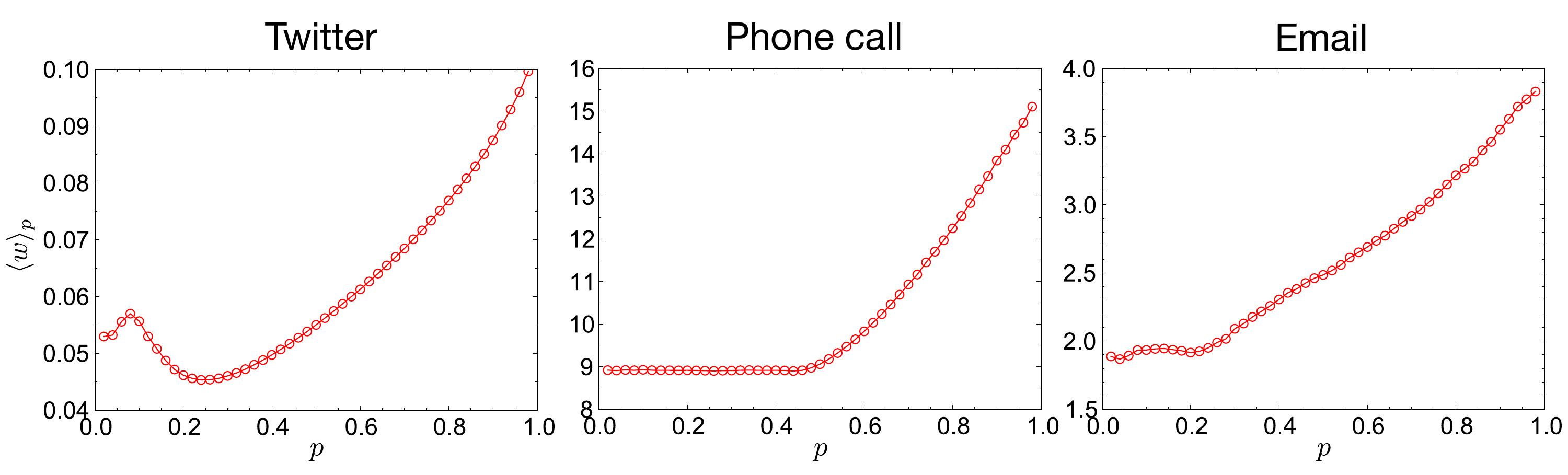}
\caption{Average weight $\langle w \rangle_p$ of the fraction $p$ of weakest ties versus $p$. Weak links have low overlap (on the left of the x axis) while strong links have high overlap (on the right).}
\label{fig:cumulative_ov_link_weight}
\end{figure}

As a first step, we aim to confirm that strong ties carry more traffic. To this end we plot the average link weight versus overlap. More precisely, following Onnela et~al.~\cite{JP2007pnas}, we define the average weight $\langle w \rangle_p$ over the fraction $p$ of weakest ties (links with lowest overlap), and plot it as a function of $p$.
As shown in Fig.~\ref{fig:cumulative_ov_link_weight}, the average link weights in the three datasets increase as a function of tie strength. Strong ties carry more traffic than weak ties, confirming that people tend to communicate more with close friends, or others with very similar social circles. The observed pattern is consistent with the weak tie hypothesis and with several previous empirical studies~\cite{friedkin1980test,JP2007pnas,onnela2007analysis,cheng2010bridgeness,grabowicz2012social,pajevic2012org}. The plateaus of the average curves for the weakest ties are due to links with zero overlap. These are quite common: 5.5\% of links in Twitter, 40\% in the cell phone network, and 23.6\% in the Enron email network connect nodes without common neighbors. 

It is important to stress the diversity of the datasets considered; they reflect the usage of communication media with different purposes, governed by different norms. Despite such differences in usage patterns, the networks corresponding to the three platforms exhibit consistent characteristics. In Twitter, the result implies that users are more likely to adopt and repost messages from neighbors with similar social circles. In the phone call network, people tend to call more frequently individuals with very similar contact lists. In the email network, people working in the same or close divisions of the corporation and thus sharing many common coworkers have more email exchanges. The emerging picture in such diverse networks provides strong evidence for the generality of the first part of Granovetter's weak tie theory.

\subsection{Attention on Weak Ties}

The second part of the weak tie hypothesis states that weak ties function as key communication channels in the social network by conveying important information that one is unlikely to discover through strong ties. Removing a strong tie is unlikely to have a significant effect on our access to information generated in our circle of friends, as alternative contacts could provide the same information. On the other hand, the removal of a weak tie could prevent us from being exposed to information from another community, to which the weak tie provides a bridge. This intuition suggests that more attention could be devoted to information received through weak ties. 

\begin{figure}
\centering
 \includegraphics[scale=0.2]{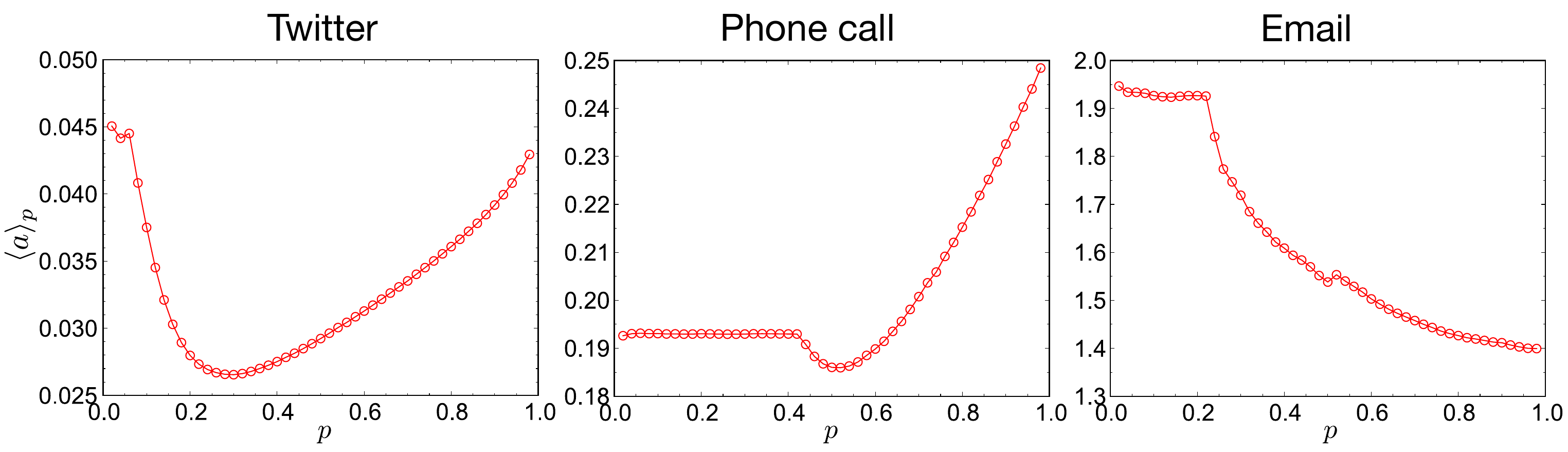}
\caption{Average link attention $\langle a \rangle_p$ of the fraction $p$ of weakest ties versus $p$. Weak ties have low overlap (on the left of the x axis) while strong ties have high overlap (on the right). The flat portion of the phone call curve for low $p$ corresponds to a high number of links with zero overlap, i.e., connecting nodes with no common neighbors.}
\label{fig:cumulative_ov_link_attention}
\end{figure}

Let us compute the average link attention $\langle a \rangle_p$ over the fraction $p$ of weakest ties (links with lowest overlap), and plot it as a function of $p$. While the three datasets show the same qualitative behavior in link weights (Fig.~\ref{fig:cumulative_ov_link_weight}), they exhibit crucial differences in the allocation of attention versus tie strength, as reported in Fig.~\ref{fig:cumulative_ov_link_attention}.

The attention curve is U-shaped in the Twitter network --- a positive correlation between attention and overlap for strong ties but a negative correlation for weak ties suggests that people are likely to allocate much attention on both very weak and very strong ties. 
The U-shape is less evident in the phone call network. Weak ties acquire attention slightly more than intermediate ties while the majority of attention is assigned to strong connections.

However, the trend is reversed in the Enron email network, where weak ties are dominant in attracting attention and there is a negative correlation between the amount of attention per tie and its strength.

A possible interpretation for the observed U-shaped attention curves in Twitter and phone data stems from two coexisting trends: on one hand, people are actively maintaining their social relationships by frequent interactions with close friends, so that strong ties capture much attention; on the other hand, people are paying attention to novel and useful information from weak ties. We can argue that a \emph{typical} user pays attention to \emph{both} weak and strong ties. Some users may pay attention to their strongest ties while others may pay attention to their weakest ties. It is conceivable that both tendencies coexist. In the aggregate, attention is split between weak and strong ties.

In Twitter, people follow close friends (strong ties) as well as other important information sources (weak ties). Hence we can observe a combined effect of the attention allocation toward both ends of the tie strength spectrum.
It seems plausible for the phone call network to be more driven by social interactions. People often call their closest friends, accounting for the greater attention toward strong ties. Calls to weak ties, such as consumer service hotlines, command attention but are much less common. 
In contrast, the email exchanges in the Enron dataset happen within a corporation and therefore we presume the network to be information-driven. The tendency for maintaining social relationships in such a network is hardly expected, consistently with the little attention observed on strong ties.
This interpretation of the attention patterns, driven by the distinction between information-driven and social-driven communication, is further explored in the next section.

\section{Social and Informational Links}

Attention concentrates on either very weak or very strong ties, as seen in Fig.~\ref{fig:cumulative_ov_link_attention}. We conjecture that this observed pattern may originate from the coexistence of two different, potentially competing, communication needs: maintaining social bonds and acquiring novel information. Let us first look into the different types of links in the three networks that might account for these two distinct tendencies.

Micro-blogging systems like Twitter, Tumblr, Weibo, and Google+ have several fundamental differences from offline social networks. These systems are designed for efficient information sharing, not only for maintaining mutual friendships. People may establish directed connections unilaterally, and therefore links do not necessarily represent relationships of mutual trust or reciprocal friendship.  Many users in micro-blogging platforms follow unknown but interesting others, such as musicians, politicians, technology experts, news sources, and brands. Owing to this special mechanism in micro-blogging systems, Huberman et~al.~\cite{huberman2008social} distinguished friends from followers based on the number of reply and mention interactions and pointed out that most traffic is conveyed by an underlying social networks of reciprocal friends.

A similar phenomenon can be found in the phone call network. Real-world friends frequently talk to each other on phone and the interactions are usually intensive, mutual, and long-lasting. Meanwhile, business hotlines and customer services get calls from individual callers on an occasional basis, and the ties between them are expected to be weak and non-mutual.

In the Enron email network, most messages are supposed to be business- or information-driven, and therefore the social activity is weaker than in the Twitter or call networks. The number of exchanges on a tie may still be dependent on how much overlap two individuals have at work, and these routine email exchanges are more likely to go through both directions. However, cross-division communication on a weak tie, though maybe not mutual (i.e., an announcement from the board), is expected to be more crucial and of higher priority, thus attracting more attention.

The social relationship between real-world friends is expected to be different from one between unknown people or coworkers (i.e., a Twitter user following a celebrity, a consumer calling a business hotline, or two coworkers with no personal contact). The former reflects existing social ties, while the latter represents information gathering. We therefore refer to these two classes of connections as \emph{social links} and \emph{informational links,} respectively. 

\begin{figure}
\centering
 \includegraphics[scale=0.2]{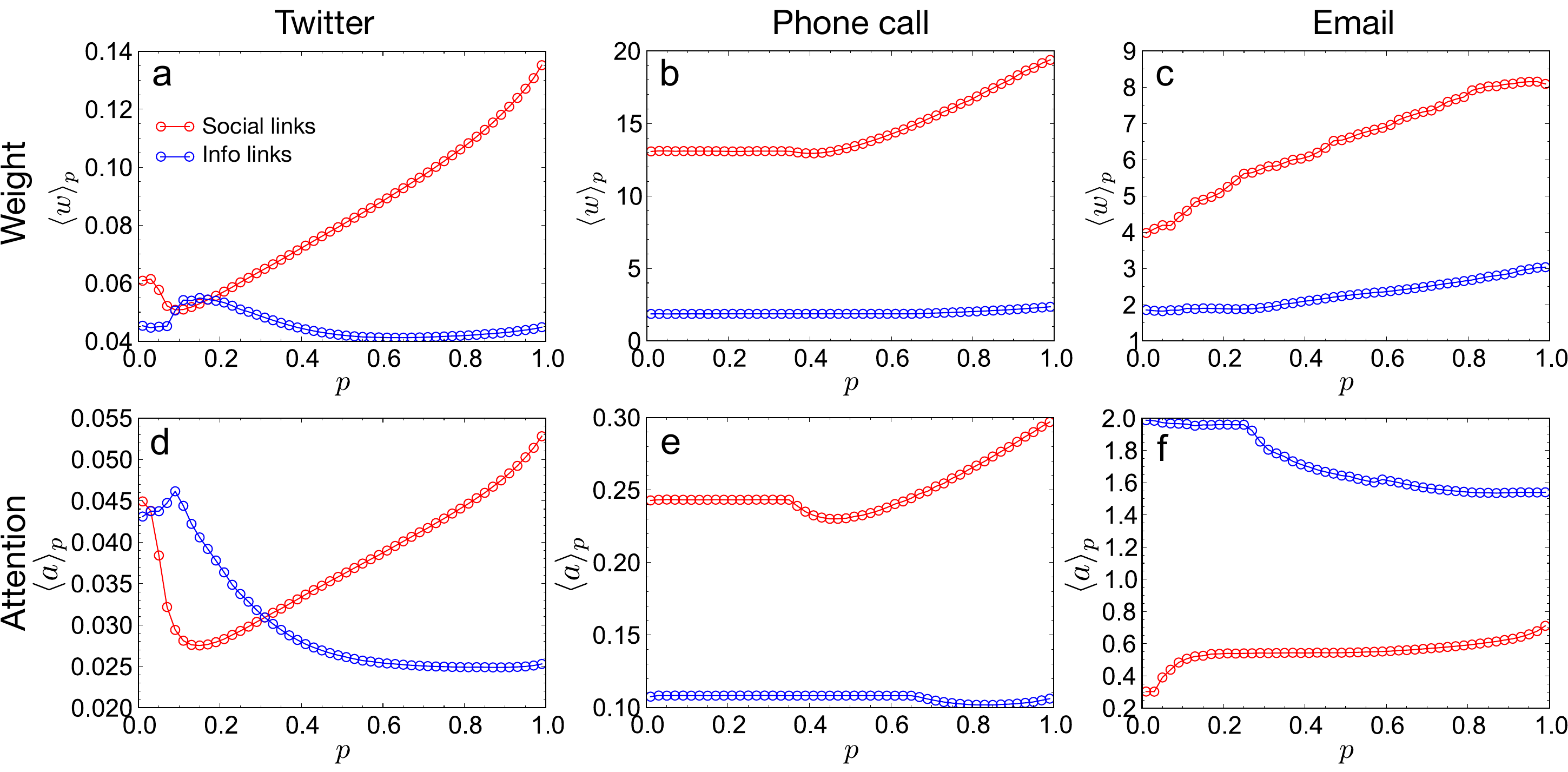}
\caption{Social links versus information links in terms of weight and attention allocation in three networks. In panels (a), (b) and (c) we plot the average link weight $\langle w \rangle_p$ of the fraction $p$ of weakest ties versus $p$. In panels (d), (e) and (f) we plot the average link attention $\langle a \rangle_p$ of the fraction $p$ of weakest ties versus $p$ }
\label{fig:social_info_links}
\end{figure}

We consider \emph{mutual} links as social and \emph{unilateral} ones (i.e., unreciprocated Twitter followers,  phone calls, and emails) as informational~\cite{huberman2008social,Granovetter,JP2007pnas,karsai2011small}. Let us compare the use of these classes of connections by separately computing average link weight and attention as a function of link overlap for social and informational links, respectively. As shown in Fig.~\ref{fig:social_info_links}, we observe clear distinctions between the two types of links in terms of the allocation of both traffic and attention. More importantly, the distinctions provide us with an interpretation of the different distributions of attention observed in the three networks (Fig.~\ref{fig:cumulative_ov_link_attention}).

Let us start with a discussion of link weights in Fig.~\ref{fig:social_info_links}(a-c). In all three networks, social links have larger weights than informational ones, irrespective of tie strengths. Their average weights increase with tie strength. The average weights of informational links, instead, do not display a robust dependence on tie strength. 
In Fig.~\ref{fig:social_info_links}(d-f) we display the attention distributions on social ties of different nature. Among \emph{social} links, strong ties attract more attention than weak ones. Among \emph{informational} links, weak ties are more appealing with regards to attention. 

Furthermore, considering that links with zero overlap play a special topological role --- a perfect bridge\footnote{Note that in our calculation, leaf nodes (with only one out-link) are removed.} connecting distant groups --- we expect to see more zero-overlap ties among informational links than among social ties. In Twitter, 7.5\% of informational links have zero overlap, compared with 4.4\% of social links; in the phone call network, about $65\%$ of informational links have zero overlap versus about $40\%$ of social links; this effect is the strongest in the email network, where 27.5\% of informational links have zero overlap as opposed to 4.1\% of social links.

The distinctions between informational and social links in terms of attention allocation help us interpret the difference between the patterns observed in Fig.~\ref{fig:cumulative_ov_link_attention}. The Twitter network allows users to maintain social contacts and information sources at the same time, and the volume of attention on social and informational links is comparable. The phone call network is more commonly used for social purposes, so informational links only win little attention overall. The email exchanges in the Enron corporate network are designed for gaining information and processing business issues, making information links dominant. 
In fact the Enron email network only contains 16\% social (mutual) links, compared to 64\% and 61\% in Twitter and phone call networks, as shown in Table~\ref{table:dataset}. 
When we aggregate attention across both classes of links (Fig.~\ref{fig:cumulative_ov_link_attention}), the increasing attention toward strong ties is explained by social interactions, while the higher attention toward weak ties originates from informational links. 
In the Twitter and phone call networks, the combined effects of the two classes of ties lead to the U-shaped attention profiles. In the email network, the predominance of informational links is consistent with the monotonically decreasing attention with increasing tie strength.

\section{Conclusion}

This chapter aimed to verify the two different aspects of the weak tie hypothesis~\cite{Granovetter} on three large empirical networks. We found that the large majority of interactions are indeed localized among strong ties. We then studied the fraction of an individual's attention directed towards a neighbor to quantify the importance of a social connection with respect to information diffusion. Interestingly we found that while strong ties do carry more traffic, weak ties succeed in attracting attention similar to or even more than strong ties. 

We hypothesize that the extent to which weak ties acquire attention can be explained by two distinct link roles, whose prevalence is network dependent. By distinguishing between social and informational links based on reciprocity, we found evidence supporting our interpretation that people interact along strong ties due to their social relationships, while looking for novel information through weak ties. In systems used for information-driven communication, such as a corporate email network, informational links are dominant, explaining higher attention toward weak ties. In systems designed for social communication, such as mobile phones, social links yield more attention and explain the importance of strong ties; however, a portion of traffic is devoted to information seeking, and so we also observe a weaker increase of attention toward weak ties. Finally, microblogs have dual social and informational purposes, explaining the non-monotonic pattern of attention versus tie strength.

Inferring the nature and purpose of a social link from its ''usage'' is challenging, but  could lead to improved ranking algorithms to prioritize social media content. This work aims to be a step in this direction.

While many studies have confirmed the first part of Granovetter's hypothesis, namely that strong ties receive more traffic in social networks, our analysis provides empirical evidence and a quantitative interpretation of the second part of Granovetter's theory, i.e., that weak ties are more important for information gathering. Until now, studies in this direction have been hampered by a lack of operational definitions of attention or importance, as well as by limits in the availability of social and communication network data that would allow one to measure these quantities at a large scale. As additional datasets of this kind become available, they will enable further refinements in our understanding of the relationships between strength, attention, and importance of social links.

\section{Acknowledgements}
  We would like thank Albert-L\'{a}szl\'{o} Barab\'{a}si for the mobile phone cell dataset used in this research, Twitter for providing public streaming data, and the Enron Email Analysis Project at UC Berkeley for cleaning up and sharing the Enron email dataset. MK acknowledges support from LABEX MiLyon. This work was partially funded by  NSF grant CCF-1101743 and the James S. McDonnell Foundation.

\bibliographystyle{unsrt} 
\bibliography{weaktie}

\end{document}